\documentstyle{article}
\title{
\bf Gonihedric 3D Ising actions
}

\author{ {\it D.A. Johnston} \\
         and \\
         {\it Ranasinghe P. K. C. Malmini $^{(a)}$}\\ \\
         Dept. of Mathematics\\
         Heriot-Watt University\\
         Riccarton\\
         Edinburgh, EH14 4AS\\
         Scotland}
\begin{document}
  \maketitle
                      {\Large
                      \begin{abstract}
%
We investigate a 
generalized Ising action containing 
nearest neighbour, next to nearest
neighbour and plaquette terms that has been suggested as a potential
string worldsheet discretization on cubic lattices 
by Savvidy and Wegner.
We use both
mean field techniques and Monte-Carlo simulations
to sketch out the phase diagram.

The Gonihedric (Savvidy-Wegner) model 
has a symmetry that allows 
any plane of spins
to be flipped with zero energy cost, which 
gives a highly degenerate vacuum state.
We choose boundary conditions
in the simulations that eliminate this degeneracy and allow the
definition of a simple ferromagnetic
order parameter.
This in turn allows us to extract the magnetic
critical exponents of the system.
\\ \\
\\ \\ \\ \\ \\ \\ \\ \\
\\ \\ \\ \\ \\ \\ \\ \\
\\ \\ \\ \\ \\
$(a)$ {\it Permanent Address:} \\
Department of Mathematics\\
University of Sri Jayewardenepura\\
Gangodawila, Sri Lanka.
%
                        \end{abstract} }
%
  \thispagestyle{empty}
%
%
  \newpage
%
                  \pagenumbering{arabic}

\section{Introduction}

In a series of recent papers \cite{1} Savvidy, Wegner and co-workers
suggested a Gonihedric random surface
action which could be written as
\begin{equation}
S = {1 \over 2} \sum_{<ij>} | \vec X_i - \vec X_j | \theta (\alpha_{ij}),
\label{e4a}
\end{equation}
on triangulated surfaces, where
$\theta(\alpha_{ij}) = | \pi - \alpha_{ij} |^{\zeta}$
with $\zeta$ some exponent
and $\alpha_{ij}$ is the angle between the
embedded neighbouring triangles with common link $<ij>$. 
This action is a robust
discretization of the linear size
of a surface, which is a well-defined geometrical
notion that may be constructed in various equivalent
ways. It was intended as an alternative to
gaussian plus extrinsic curvature lattice actions
of the form
\begin{equation}
S = \sum_{<ij>} ( \vec X_i - \vec X_j )^2 +
\lambda \sum_{\Delta_i, \Delta_j} ( 1 - \vec n_i  \cdot \vec n_j )
\label{e01}
\end{equation}
which have been much explored \cite{2} as discretizations
of rigid membranes and strings \cite{3}. 
Although a simulation showed that the action of equ.(\ref{e4a})
produced flat surfaces \cite{4}, potential problems
arising from the failure to suppress the wanderings of vertices
in the plane were pointed out in \cite{5}
for the action with $\zeta=1$. Possible ways to cure
this are to add additional Gaussian or linear terms to the action
\cite{5a}
or, more satisfactorily, to simply choose $\zeta<1$.

A study of the scaling of the string tension and mass gap
in a dynamically triangulated model with
an additional linear term produced inconclusive results \cite{6},
as have simulations of the gaussian plus extrinsic curvature
action, because of the difficulties of simulating a 
complicated action on a dynamical surface.
There is thus some incentive to investigate
alternative approaches to such surface models where
the computational costs are less onerous.
One such approach
for regularizing the Gonihedric action is to
restrict the allowed surfaces to a (hyper)cubic lattice. This 
has been pursued in some detail analytically in \cite{7,8,8a}
and one numerical simulation carried out in three dimensions
\cite{8}.
Recently Pietig and Wegner \cite{8b} have demonstrated
with a Peierls contour argument that a transition 
{\it does} exist in the three-dimensional case and 
obtained a bound on the critical temperature
that is not contradicted by the simulations.
The crucial observation 
for this work is that the surface theory
on a cubic or hypercubic lattice can be written equivalently as a
generalized Ising action, where the boundaries
between the spin clusters are the surfaces of the original model.
The latticized Gonihedric model assigns a non-zero action
only to right angled bends in the surface and self-intersections.
Normalizing the weight for a right-angled bend on a link
appropriately leaves
a free parameter $\kappa$ that gives
the relative weight
of a self-intersection of the surface on a link. 
The energy of a surface on a cubic
lattice is thus given explicitly by
$E=n_2 + 4 \kappa n_4$, where $n_2$ is the number
of links where two plaquettes meet at a right angle
and $n_4$ is the number of links where four plaquettes
meet at right angles.

A Hamiltonian which reproduces
this energy on a cubic lattice has the form
\begin{equation}
H= 2 \kappa \sum_{<ij>}^{ }\sigma_{i} \sigma_{j} -
\frac{\kappa}{2}\sum_{<<i,j>>}^{ }\sigma_{i} \sigma_{j}+
\frac{1-\kappa}{2}\sum_{[i,j,k,l]}^{ }\sigma_{i} \sigma_{j}\sigma_{k} \sigma_{l}
\label{e1}  
\end{equation}
which is of generalized Ising form and contains nearest neighbour ($<i,j>$),
next to nearest neighbour ($<<i,j>>$) and round a plaquette ($[i,j,k,l]$)
terms. 
Such actions are not new, having been investigated in some detail using
both mean field methods and simulations in \cite{9}. 
However,
the particular
combination of coefficients arising in equ.(\ref{e1}) was not considered 
explicitly in this work because it corresponds
to a particularly degenerate set of couplings.
This degeneracy manifests itself in an extended symmetry
in the model:
for any value of $\kappa$
it is possible to flip a plane of spins with no energy penalty,
providing the flipped plane does not intersect any existing
portion of surface.
This gives a vacuum degeneracy reminiscent of a gauge theory,
the difference with a true gauge theory
being that here the symmetries are quasi-global, giving only
the freedom to flip entire planes rather than local spins.
Related surface models have also been
simulated directly in \cite{9a},
but again the set of coefficients
appearing in equ.(3)
was not explored in this work.
A very rich phase structure was observed in \cite{9}, in common with 
other Ising models with extended interactions \cite{10}
of various sorts which display first and second order phase boundaries
as well as incommensurate phases.
Given the
generic richness of the phase diagrams
for such generalized Ising models
and the additional symmetries
present in the Gonihedric model, the action of equ.(\ref{e1}) merits investigation
from purely statistical mechanical considerations as well
as from the point of view of finding potential continuum string theories.

In the context of string theory one is looking for 
a continuous transition (or transitions) at which
a sensible continuum surface theory may be defined.
It is perhaps worth recalling that 
even this does not guarantee
a good continuum surface theory. The interfaces
in the standard nearest neighbour Ising model in
three dimensions, 
which has a continuous phase transition,
have been investigated in some detail
recently and found to be very porous objects,
decorated with lots of handles at the scale of the lattice
cutoff \cite{11}. Ideally one might hope that
the surfaces generated by the Gonihedric action were
smoother, given that it is derived from a 
sort of stiffness term.

Our motivation in this paper is to investigate
the action of equ.(\ref{e1}) in order to sketch out the
gross features
of its phase structure.
We concentrate on $\kappa=1$ as in \cite{8}, but also discuss other values.
A few
cautionary words are in order before we go on to discuss
the mean field approach and simulations. 
As we have noted 
the ground state of the action in
equ.(\ref{e1}) is very degenerate as 
planes of spins parallel
to any of the cube axes can be flipped at no energy cost. 
In the case $\kappa=0$ the degeneracy is even larger
as diagonal planes may be flipped now also. 
The ability to flip arbitrary spin planes makes defining a magnetic
order parameter rather
problematic. Even the staggered local order parameters defined
in \cite{9} would miss the lamellar phases with arbitrary
intersheet spacings that could be generated
at no cost by flips of spin planes. 

The simulations for $\kappa=1$ in \cite{8} were restricted
to measuring the energy and specific heat as
the exhaustive global order parameters suggested there
\begin{equation}
M^{\mu} = \left< {1 \over L^3} \sum_i \sigma_i^{\mu} (vac) \; \sigma_i \right>
\end{equation}
(where $\sigma_i^{\mu} (vac)$
is one of the possible vacuum 
spin configurations with $\mu = 1,2 \ldots 2^{3L}$ for $\kappa=0$ 
or $\mu = 1,2 \ldots 3 \times 2^L$ for $\kappa \ne 0$ on a lattice of size $L$)
would have been prohibitively slow to measure on even moderately
sized lattices. It is possible to do rather better, however, by making
use of the freedom in choosing boundary conditions on a finite lattice.
It is customary to employ periodic boundary conditions in attempting
to extract critical exponents from a simulation as these tend to
minimize the finite size effects. It is clear that {\it fixed} 
boundary conditions in the Gonihedric model would penalize
flipped spin planes by at least a perimeter energy, at the
possible expense of greater finite size effects. A quick test
simulation shows that such boundary conditions do
pick out the purely ferromagnetic ground state from the
many equivalent possibilities and allow the measurement
of the simple ferromagnetic order parameter
\begin{equation}
M = \left< {1 \over L^3} \sum_i \sigma_i \right>.
\end{equation}
One can, in fact, have the best of both worlds by continuing
to employ periodic boundary conditions to reduce the finite size
effects whilst fixing any two perpendicular planes
of spins to pick out the ferromagnetic ground state.
The fixed planes of spin are, in effect, more akin
to a gauge fixing of the spin flip symmetry
than boundary conditions {\it per se}. 
In the simulations reported later in this paper
we employed three fixed perpendicular planes  
of spins as a safety measure, with essentially 
identical results.

\section{Zero Temperature and Mean Field}

As the Gonihedric model is
a special case of the general action considered in \cite{9}
we can apply the methods used there
for both the zero temperature phase diagram and mean field 
theory. 
For the zero temperature 
case this involves writing the full lattice
Hamiltonian as a sum over individual cube Hamiltonians
\begin{equation}
h_c = \frac{\kappa}{2}\sum_{<i,j>} \sigma_{i} \sigma_{j} - \frac{\kappa}{4}  \sum_{<<i,j>> }\sigma_{i} \sigma_{j} 
+  \frac{1-\kappa}{4} \sum_{[i,j,k,l]}\sigma_{i} \sigma_{j} \sigma_{k} \sigma_{l}
\end{equation}
and observing that if the lattice can be tiled by
a cube configuration minimizing the individual $h_c$
then the ground state energy density is
$\epsilon_0 = min\;  h_c$. 

We list the inequivalent spin
configurations on a single cube and their
multiplicities in Table.1 using the same notation 
as \cite{9} but with our choice of couplings
to highlight the degeneracies that appear with the Gonihedric
action.
In the list of spins the first column represents one face of the cube
and the second the other.
In the table two configurations are considered equivalent if one can be transformed
into the other by reflections and rotations
or if they are related by a global spin flip. The antiferromagnetic image
of a configuration
is obtained by flipping the three nearest neighbours and the spin
at the other end of the cube diagonal from a given spin and is denoted by
an overbar.
With the Gonihedric values of the couplings the freedom to flip spin planes is clear
even at this level as $\psi_0$, which would represent a ferromagnetic state
when used to tile the lattice, and $\psi_6$ which would represent flipped
spin layers,
have the same energy for any value of $\kappa$. The higher energy configurations
$\psi_4$ and $\psi_{\bar 4}$ are also identical. 
The degeneracies increase when
$\kappa=0$, the club
of
states of energy $-3/2$ is now composed of $\psi_0, \psi_{\bar 0}, \psi_6, \psi_{\bar 6}$
and various extra degeneracies appear for higher
energy states. The new ground states $\psi_{\bar 0}, \psi_{\bar 6}$
reflect the additional freedom to flip diagonal planes of spins that is present at
$\kappa=0$.

In the mean field approximation the spins
are in effect replaced by the average site magnetizations.
The calculation 
of the mean field free energy is an elaboration of the
method used
above to investigate the ground states
in which the energy is decomposed into a sum of individual cube terms.
The next to nearest neighbour and plaquette interactions
in the Gonihedric model give
the total mean field 
free energy as the sum of elementary cube free energies $\phi(m_{c})$, given by
\begin{equation}
\phi{(m_{C})}=- \frac{\kappa}{2}\sum_{<i,j>\subset C} m_{i} m_{j} + \frac{\kappa}{4}  \sum_{<<i,j>>\subset C }m_{i} m_{j} \] 
\[    -  \frac{1-\kappa}{4} \sum_{[i,j,k,l] \subset C}m_{i} m_{j} m_{k} m_{l} + \frac{1}{16}
\sum_{i \subset C}[(1+m_{i})ln(1+m_{i})+ (1-m_{i})ln(1-m_{i})]
\end{equation} 
where $m_{C}$ is the set of the eight magnetizations of the elementary cube.
This gives a set of eight mean-field equations
\begin{equation}
\frac{\partial\phi(m_{C})}{ \partial m_{i}}_{(i=1 {\ldots} 8)} =0
\end{equation}
(one for each corner of the cube)
rather than the familiar single equation for the standard nearest neighbour Ising
action.
More explicitly, we have
\begin{eqnarray}
m_{1}&=& tanh[4\beta  \kappa(m_{2}+m_{4}+m_{5})-2\beta \kappa(m_{3}+m_{6}+m_{8}) \nonumber \\
     & & + 2\beta(1- \kappa)(m_{2}m_{3}m_{4}+ m_{2}m_{5}m_{6}+m_{4}m_{5}m_{8}) ] \nonumber \\
     &  \vdots& \nonumber \\
m_{8}&=& tanh[4\beta \kappa(m_{2}+m_{5}+m_{7})-2\beta \kappa(m_{1}+m_{3}+m_{6}) \nonumber \\
     & & + 2\beta(1-\kappa)(m_{3}m_{4}m_{7}+m_{1}m_{4}m_{5}+m_{5}m_{6}m_{7})] 
\label{e2a}
\end{eqnarray}  
where we have labelled the magnetizations on a face of the cube counterclockwise $1 \ldots 4$ 
and similarly for the opposing face $5 \ldots 8$
as shown in Figure.1.
If we solve these equations iteratively we arrive at 
zeroes for a paramagnetic phase or various combinations
of $\pm 1$ for the magnetized phases on the 
eight cube vertices, and the mean field ground state
is then give by gluing together the elementary cubes consistently
to tile the complete lattice, in the manner
of the ground state discussion. 

Turning loose a numerical solver on the mean field equs.(\ref{e2a}) gives generically a 
single transition to one of the phases
listed in Table.1 from the high temperature
paramagnetic phase. The transition temperatures and the resulting low temperature
phase are listed in Table.2. 
We have taken the liberty of carrying out global
flips where necessary to tidy up the table.
Rather remarkably, we see that apart from $\kappa=0$
the transition appears to be to the simple ferromagnetic phase, $\psi_0$. However, remembering
that $\psi_0$ and $\psi_6$ have the same energy the best we can say is that
we end up in a layered phase with arbitrary interlayer spacing in all directions.
Although the $\kappa=0$ case appears to be superficially different,
the $\psi_{\bar 0}$ phase that is found at low temperature here
is one of the phases that is degenerate with $\psi_0$ and $\psi_6$
when $\kappa=0$.
Although $\kappa=1$ fits the pattern as far as
a transition to $\psi_{0,6}$ at decreasing $\beta$
is concerned it appears to be rather atypical in that
further transitions are observed at larger $\beta$.
However, this is a numerical instability
that is peculiar to this particular value of $\kappa$.
It was observed in \cite{9} that an iterative solution
of the mean field equations
written in the form 
\begin{equation}
m_i^{(n+1)} = f[E_{,i} (m^n)]
\end{equation}
where $E$ is the individual cube Hamiltonian
could fail to converge if an eigenvalue of
$ \partial m^{(n+1)}_i / \partial m^n_j$
was less than $-1$. Modifying the equations
to 
\begin{equation}
m_i^{(n+1)} =  { \left( f[E_{,i} (m^n)] + \alpha m^n_i \right) \over 1 + \alpha}
\end{equation}
for suitable $\alpha$ cures this. This is precisely what happens
for $\kappa=1$, where introducing a non-zero $\alpha$ suppresses the
extra ``transitions''.

In summary, the mean field theory suggests a rather simple
phase diagram for the Gonihedric model
with action equ.(\ref{e1}), with a single transition
from a paramagnetic phase
to a degenerate ``layered'' phase
that is pushed down to $\beta=0$ at large $\kappa$.
The low temperature phase is generically
of the $\psi_{0,6}$ type, apart from $\kappa=0$ where
we see a $\psi_{\bar 0, \bar 6}$ phase that is degenerate
with these. The degeneracy of the
ground states that are indicated by these results
are, as they should be,
consistent with the symmetries of the original
full action.
We now go on 
to see how the zero-temperature and mean field results 
tally with a direct Monte-Carlo
simulation.

\section{Simulations}

We carried out simulations 
with $\kappa=1$ on lattices
of size $10^3,12^3,15^3,18^3,20^3$ and $25^3$
and for $\kappa=0,2,5,10$
on lattices of size $10^3,15^3,20^3$ and $25^3$.
Unless stated otherwise periodic boundary conditions were imposed
in the three directions and three internal perpendicular planes
of spins fixed to be $+ 1$.
We carried out 50K 
sweeps for most $\beta$ values,
increasing to 500K sweeps near the observed
phase transition point. Measurements were carried out
every sweep after allowing sufficient time 
for thermalization.
A simple Metropolis update was used because of the difficulty 
in concocting a cluster algorithm for a Hamiltonian with
such complicated interaction terms. The program was tested
on the standard nearest neighbour Ising model and 
the some of the parameters used in 
the generalized Ising models of \cite{9} to ensure it was working.

We measured the usual thermodynamic quantities for the model:
the energy $E$, specific heat $C$, 
(standard) magnetization $M$, susceptibility $\chi$
and various cumulants. As the large $\beta$ limit of the
energy should be determined by the zero-temperature
analysis of the preceding section, we consider
the energy first.
The absolute value of the energy for various $\kappa$ 
on lattices of size $L=20$ is plotted against $\beta$
in Figure.2, where we can see that the zero temperature
prediction of $3(1+ \kappa) / 2$ is satisfied 
with good accuracy for sufficiently large $\beta$.
We can therefore observe that the zero-temperature/mean-field
analysis has correctly identified the ground state(s) of the theory:
$\psi_{0,6}$ for $\kappa \ne 0$; or $\psi_{0,\bar 0,6,\bar 6}$ for
$\kappa=0$ as these are the only states with the observed
energies. Having satisfied ourselves that the 
simulation is finding the correct ground state energy, we 
can go on to consider extracting some of the critical exponents for the transition.
In what follows we will, as advertized, discuss in some detail the case $\kappa=1$
before commenting more briefly on the other
values of $\kappa$. 
 
With only periodic 
boundary conditions the possibility of the simple ferromagnetic
ordered state $\psi_0$ can be excluded by looking at
the magnetization $M$, which for all
$\kappa$ is either zero or fluctuates wildly as $\beta$
is changed, reflecting the freedom to flip spin planes.
Curtailing this freedom by fixing the three perpendicular
spin planes picks out the transition
to a simple ferromagnetic ground state
and the low temperature limit
of the magnetization becomes
one for all $\kappa$.
The magnetization cumulant
\begin{equation}
U_M =  { <M^4> \over <M^2>^2 }
\end{equation}
is well defined once the three spin planes are fixed,
but it does not show the standard behaviour
of a low $\beta$ limit of three and a high
$\beta$ limit of one, asymptoting to a value slightly
larger than one at low $\beta$, as can be seen in Figure.3.
This is because the fixed planes still leave a residual
magnetization at low $\beta$, which is sufficient 
to make this limit look magnetized for the sizes
of lattice we simulate. Nonetheless, it is still
possible to apply the usual scaling analysis in the critical region,
and the crossing of the cumulant plots for different latttice
sizes gives an estimate of $\beta_c = 0.44(1)$ for the critical
temperature, which is in good agreement with the value
reported in \cite{8} that was extracted by looking
at the change in behaviour of the spin/spin correlation
as $\beta_c$ was approached. As noted in \cite{8}
this $\beta_c$ is very close (in our case
within the error bars) to that of the standard
two-dimensional Ising model on a square lattice. 

The rather sharp nature of the crossing, or more accurately
collapse down to a single line, makes it difficult to extract
a value for $\nu$ from the ratio of the slopes 
of the Binder's cumulant curves at the
critical point, so we take a different tack and 
consider the scaling of the maximum slope,
which we would also expect to scale as
$L^{1 / \nu}$ \footnote{This tactic
works well in, for instance, simulations of Ising
models coupled to two-dimensional quantum gravity.}.
This gives an estimate of $\nu = 1.2(1)$.
We can also look at both
the finite size scaling and direct fits to 
the susceptibility $\chi$, namely
$\chi = A L^{\gamma \over \nu}$ and $\chi = \tilde A | \beta - \beta_c | ^{\gamma}$,
to attempt to extract $\nu$. We choose this in preference to the
specific heat fits $C = B + D L^{\alpha \over \nu}$ and $C = \tilde B + \tilde D | \beta - \beta_c |^{\alpha}$ 
because of the absence of an adjustable constant.
The susceptibility data is plotted in Figure.4, and shows a clear peak. This should be contrasted
with the case of no fixed planes where the $\beta>\beta_c$ region is rendered meaningless noise by
the lack of a well-defined magnetization for the myriad of ground states.
We find the finite size scaling fit gives $\gamma / \nu = 1.79(4)$ with
a high quality, and feeding the critical value of $\beta_c=0.44$ into the direct fit
on the $L=25$ lattice gives $\gamma=1.60(2)$ with rather lower quality. The deduced
value for $\nu$ is thus $1.10(5)$. These fits give values
close to those for the standard two-dimensional Ising model
with only nearest neighbour interactions, where we have $\gamma=1.75, \nu =1$.

As a consistency check on our values of $\beta_c$ and $\nu$, we plot 
the $\beta$ values where the specific heat peaks and the $\beta$ values 
where the maximum slope of the Binder's cumulant curve occurs, 
both of which can serve as estimates of the pseudocritical
temperature on finite size lattices, against
$L^{-{1 \over \nu}}$. We would expect this to be a straight line with
intercept $\beta_c$. The plot is shown in Figure.5 for the 
choice $\nu=1$ with other 
values in this neighbourhood giving essentially identical results. 
We can see that the estimate of $\beta_c=0.437(7)$
coming from the two possibilities is both self-consistent and in agreement
with the value obtained from cumulant crossing.

The above, apparently consistent, set of results presents us with something of a
dilemma when it comes to the analysis of the specific heat peak, which is shown in
Figure.6. The hyperscaling relation $\alpha = 2 - \nu d$ indicates that, if a 
value of $\nu \simeq 1$ is to be believed, the specific heat should display a 
cusp ($\alpha = 2 - \nu d \simeq -1$) rather than a divergence. This does not appear
to be the case for the data in the figure. 
Setting aside the hyperscaling result for the moment
and performing direct power law fits to 
$C = B + D L^{\alpha \over \nu}$ and 
$C = \tilde B + \tilde D | \beta - \beta_c |^{\alpha}$ gives poor fits.
A logarithmic fit of the form
$C = B + D \log ( L )$ or $C = \tilde B + \tilde D \log (\beta - \beta_c)$
gives much better, but still not particularly good, results
so the evidence is inconclusive.

Another line of attack
for obtaining an estimate of $\alpha$ is to use the finite size
scaling of the energy itself
$E \simeq E_0 + E_1 L^{\alpha -1 \over \nu}$. With
direct measurements in this form one gains nothing
in general over the specific heat 
fits as there is still a regular term $E_0$
to be dealt with, although for models with higher than
second order transitions this approach may be preferable \cite{wh}.
However, if one has measurements for two different sets
of boundary conditions available the regular term would be
expected to be the same for both and the energy 
difference could be used for a simple power law fit
to extract $( \alpha -1 ) /  \nu$. We are currently
measuring the string tension in the Savvidy model using
two sets of fixed boundary conditions \cite{esp}, one with
all positive spins and one with half positive and half negative
spins \footnote{It is not possible to use antiperiodic
boundary conditions in the Savvidy model to enforce
an interface because of the plane flip symmetry. Something
more coercive, in the form of these fixed boundary conditions is
required.}. For these measurements we would expect
\begin{equation}
\Delta E = E_{++} - E_{+-} = A L^{\alpha - 1 \over \nu}
\end{equation}
where $E_{++}$ is the energy for all positive spins and
$E_{+-}$ is the energy for half positive and half negative
spins.
A fit gives $( \alpha - 1 ) / \nu = -1.3(2)$, which is still
marginally consistent with $\alpha=0$.

There are two possible interpretations of the results. The first is that
the value of $\nu$ measured is simply wrong, not inconceivable as it 
appears either as a derived quantity from 
the slope of the cumulant or from the two fits 
to the susceptibility rather than being
measured directly.
However, a second possible interpretation
is to accept the fits to $\nu$ at face value and posit
that the model sees an effective dimension of $d=2$
in order to avoid violating the hyperscaling relation.
In this case we could recover the full set of 
two-dimensional Ising model exponents.
Although this would be highly unusual, it should be remembered
that the energy in the model is essentially linear in form
rather than being an area, so there is some resemblance to
a two-dimensional model.

A direct fit to the magnetization exponent on the largest
lattice size $M \simeq |\beta - \beta_c|^{\beta}$
(with apologies for the profusion of betas!) with $\beta_c$
fixed to be $0.44$ gives $\beta = 0.12(1)$, but the quality
is low, whereas a finite size scaling fit gives a
much lower value of $\beta / \nu = 0.04(1)$. It is
possible that the fixed spin planes, whose residual
magnetization we have not allowed for in the fits,
are biasing the finite size scaling fit, but intuitively
one would expect their
effects (of order $3 / L$) to increase 
rather than decrease the estimated exponent
by pushing up the measured magnetization on the smaller lattices.

We now discuss more briefly the other $\kappa$ values
that were simulated, namely $\kappa=0,2,5,10$. 
Firstly we can note that in qualitative
terms the transition appears similar to the $\kappa=1$ case,
a not entirely trivial result as $\kappa \neq 1$ introduces
round a plaquette interactions that are missing for $\kappa=1$.
With the periodic plus fixed plane boundary conditions
we still have peaks in the susceptibility and specific heat
and a ferromagnetically ordered phase appearing 
at low temperature.
The mean field
analysis suggests that as $\kappa$ is increased
$\beta_c$ should drop sharply. This is {\it not} observed
in the simulations, the crossing of the Binder's
cumulants indicating no change within the error
bars for the estimates of $\beta_c$,
giving $\beta_c = 0.44(1)$
from $\kappa=1$ to $\kappa=10$. 
There is a sharper difference with the $\kappa=0$ results
which show a crossing at $\beta_c = 0.50(1)$.
In general
mean field theory will underestimate $\beta_c$, so the 
measured results are in agreement with this
and do not contradict the bound obtained in
\cite{8b}.

The similarity of the transitions
for different $\kappa$ extends beyond $\beta_c$. The
measurements of $\gamma / \nu$ listed in Table.3 below for
all the non-zero $\kappa$ 
suggest that the critical behaviour is unchanged
by varying $\kappa$.

\begin{center}
\begin{tabular}{|c|c|c|c|c|} \hline
$\kappa$   & $1$    & $2$    & $5$ & $10$     \\[.05in]
\hline
$\gamma / \nu$  &  $1.79(4)$ & $1.6(1)$ &  $1.9(1)$ & $1.75(6)$   \\[.05in]
\hline
\end{tabular}
\end{center}
\vspace{.1in}
\centerline{Table 3: Fits to $\gamma / \nu$ for the non-zero $\kappa$ values.}

\noindent
The specific heat curves and magnetization present a similar story,
with all looking similar to the $\kappa=1$ case.
From this evidence it would seem that varying $\kappa$,
at least for $\kappa \ge 1$, 
gives little if any change in the exponents and transition
temperature. 

The story is slightly different for $\kappa=0$.
As we have already noted $\beta_c=0.50(1)$,
and other differences are apparent.
Without the fixed spin planes (ie with 
only periodic boundary conditions) the
susceptibility when $\beta<\beta_c$ for 
non-zero $\kappa$ values is similar to the fixed plane
case described above and becomes meaningless
when $\beta>\beta_c$ where the magnetization is ill-defined.
The $\kappa=0$ model presents qualitatively different behaviour
in that the susceptibility is one for 
$\beta<\beta_c (\simeq 0.5)$ and zero for $\beta>\beta_c$. 
However, this step 
function behaviour disappears when the fixed
spin planes are introduced and we recover a divergent peak
as for the other $\kappa$ values. The phase structure for $\kappa=0$ 
with the fixed spin planes
also appears to be broadly similar to other $\kappa$, giving
a single transition to a low temperature magnetized phase.
It would be interesting to examine in detail values of
$\kappa$ between zero and one to see if there was a
smooth change in, for example, $\beta_c$ 
as $\kappa \rightarrow 0$. This would give some indication 
of whether $\kappa=0$ really was a special point,
or joined on smoothly to the continuum of non-zero values.
A test simulation at $\kappa=0.5$ still gives
very similar results to $\kappa=1$, for example.

\section{Conclusions}

We have conducted zero-temperature, mean-field and Monte-Carlo
investigations of the generalized Ising model action suggested
in \cite{7,8,8a} as a cubic lattice discretization
of the Gonihedric string action \cite{1}
using essentially the methods of \cite{9}. Although the 
phase structure of such generalized Ising models is
generically very rich \cite{10}, the one parameter family
of models examined here seems to be a fairly
simple ``slice'' of the phase diagram, with one transition
to a layered ground state
when periodic boundary conditions are imposed. 
This degenerate layered state is a consequence of
the plane spin flip symmetry that is present in the model
for all $\kappa$, but a judicious choice of boundary conditions
- fixing enough perpendicular spin planes - allowed us
to pick out an equivalent ferromagnetic ground state
for the purposes of simulations.
The zero-temperature/mean-field analyses are in agreement
with the Monte-Carlo simulations on the nature of the ground
state and its energy, but the simulations
indicate a transition temperature that changes little,
if at all, from its value at $\kappa=1$
($\beta_c \simeq 0.44$) for other non-zero $\kappa$ values.
The mean field theory on the other hand  gives
a fairly sharp decline in $\beta_c$ as $\kappa$ is increased.

The simulations at $\kappa=1$ 
indicate that the fitted exponents,
with the exception of the finite size scaling fit to
$\beta / \nu$, 
and even the critical temperature are all in
the vicinity of those for the two-dimensional Ising model,
though given our modest statistics 
it would be foolhardy to claim they were identical on
the basis of the current fits.
Comparison with the other non-zero $\kappa$ values
also gives similar exponents and critical temperatures. 
There is some evidence
that the $\kappa=0$ model is a special case: -
in the zero temperature and mean field analyses
more ground states are allowed and in the simulations
a different critical temperature is observed
and the behaviour of the susceptibility is radically different
when no spin planes are fixed.

An immediate extension of the current work, given the closeness
of the fitted exponents to the two-dimensional Ising model,
is to carry out a higher statistics simulation
near the transition point in order to pin down
the various exponents and $\beta_c$ more accurately. 
A further test of the critical behaviour of the model
would be to investigate the scaling of the string tension
as one approached the critical point, along the lines of \cite{esp}.
The various higher dimensional
generalizations that were formulated in \cite{7,8,8a}
also merit investigation.

If we return to our original stringy motivation
it would be useful to characterize the surfaces generated
by the Gonihedric action in the style of \cite{11}
to see whether they were any less ``spongy'' than those
in the standard 3D Ising model.
As a playground for exploring plaquette discretizations
of string and gravity inspired models the generalized
Ising models clearly have some interesting quirks 
that are worthy of further exploration. It would certainly
be amusing to show that a candidate discretized string model
was a close relation of the {\it two}-dimensional
Ising model.

\section{Acknowledgements}

R.P.K.C. Malmini was supported by Commonwealth Scholarship 
SR0014.

\vfill
\eject

\vfill\eject
\centerline{\bf TABLES}
\medskip
\centerline{\bf Table.1}
\begin{center}
\begin{tabular}{|c|c|c|c|c|}    \hline
State & Top  & Bottom& Energy & Multiplicity  \\ \hline
$\psi_0$ & {\tiny + +}& {\tiny + +} & $-3/2 -3 \kappa /2$ & 2 \\
       & {\tiny + +}& {\tiny + +} & &        \\  \hline
$\psi_{\bar 0}$ & - {\tiny +}& {\tiny +} - & $-3/2 +21\kappa /2$ & 2 \\
       & {\tiny +} - & - {\tiny +}& &   \\ \hline
$\psi_1$ & {\tiny + +} & {\tiny + +} & $-3 \kappa /2$ & 16 \\
         & - {\tiny +} & {\tiny + +} & & \\  \hline
$\psi_{\bar 1}$ & - {\tiny +} & {\tiny + +} & $9 \kappa /2$ & 16 \\
         & {\tiny +} - & - {\tiny +} & &  \\  \hline
$\psi_{2, \bar 2}$ & - {\tiny +} & {\tiny + +} & $1/2  + \kappa /2$ & 24 \\
         & {\tiny +} - & {\tiny + +} & &  \\  \hline
$\psi_{3}$ & {\tiny + +} & {\tiny + +} & $-1/2 -3 \kappa /2$ & 24 \\
         &  - - & {\tiny + +} & &  \\  \hline
$\psi_{\bar 3}$ &  - -  & {\tiny + +} & $-1/2 +5 \kappa /2$ & 24 \\
         &  - {\tiny +} & {\tiny + } - & &  \\  \hline
$\psi_{4}$ &  - {\tiny +}  & {\tiny + +} & $3/2 -3 \kappa /2$ & 8 \\
         &  {\tiny + +} & {\tiny + } - & &  \\  \hline
$\psi_{\bar 4}$ &  - - & - {\tiny +} & $3/2 -3 \kappa /2$ & 8 \\
         &  - {\tiny  +} & {\tiny + + }  & &  \\  \hline
$\psi_{5}$ &  - - &  {\tiny + +} & $ -3 \kappa /2$ & 48 \\
         &  - {\tiny  +} & {\tiny + + }  & &  \\  \hline
$\psi_{\bar 5}$ &  - {\tiny  +} &  - {\tiny  +} & $  \kappa /2$ & 48 \\
         &  {\tiny  +} - & {\tiny + + }  & &  \\  \hline
$\psi_{6}$ &  - - &  {\tiny +  +} & $  -3/2 -3 \kappa /2$ & 6 \\
         &  - -   & {\tiny + + }  & &  \\  \hline
$\psi_{\bar 6}$ &  {\tiny + + } &  - -  & $  -3/2 +5 \kappa /2$ & 6 \\
         &  - -   & {\tiny + + }  & &  \\  \hline
$\psi_{7, \bar 7}$ &  - -  &  {\tiny +}  -  & $  1/2 -3 \kappa /2$ & 24 \\
         &  - {\tiny +}   & {\tiny + + }  & &  \\  \hline
\end{tabular}
\end{center}
\leftline{Table.1: The inequivalent spin configurations of a single cube
and the associated energies and degeneracies}

\vfill\eject
\medskip
\centerline{\bf Table.2}
\begin{center}
\begin{tabular}{|c|c|c|c|}    \hline
 
$\kappa$         &      $\beta_c $        &     &    \\ \hline
 
0.0       &      0.325             &  {\tiny +} - &   -  {\tiny +}       \\
          &                        &  -  {\tiny +} &   {\tiny +}  -  \\ \hline
  
0.25      &      0.31              &  {\tiny + +} & {\tiny + +}        \\
          &                        &  {\tiny + +} & {\tiny + +}         \\  \hline
0.5       &      0.278             &  {\tiny + +} & {\tiny + +}        \\
          &                        &  {\tiny + +} & {\tiny + +}         \\  \hline
1.0       &      0.167             &   {\tiny + +} &  {\tiny + +}     \\  
          &                        &   {\tiny + +} & {\tiny + +}      \\  \hline
2.0       &      0.09               &  {\tiny + +}& {\tiny + +}        \\ 
          &                        &  {\tiny + +}& {\tiny + +}         \\  \hline
5.0       &      0.0335            &  {\tiny + +}& {\tiny + +}        \\ 
          &                        &  {\tiny + +}& {\tiny + +}         \\  \hline
10.0      &      0.02              &  {\tiny + +}& {\tiny + +}        \\ 
          &                        &  {\tiny + +}& {\tiny + +}         \\  \hline
15.0      &      $<$0.02           &  {\tiny + +}& {\tiny + +}      \\
          &                        &  {\tiny + +}& {\tiny + +}   \\   \hline
\end{tabular}
\end{center}
\centerline{Table.2: The ground state configurations and transition temperatures
for various $\kappa$.}
\centerline{ The states shown appear above the quoted temperature.}
\clearpage \newpage
\begin{figure}[htb]
\vskip 20.0truecm
\includegraphics{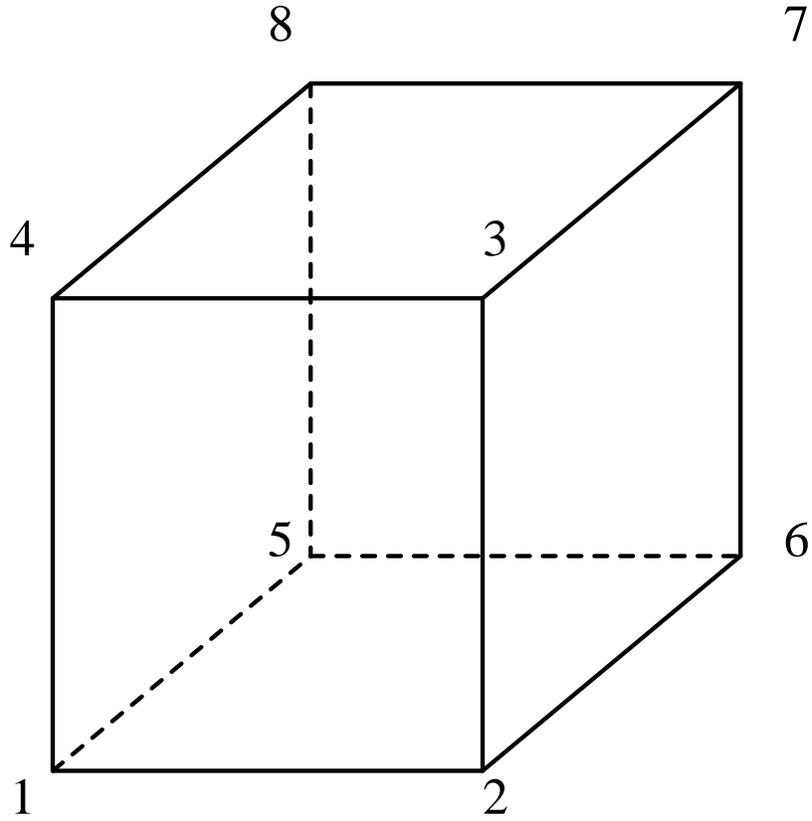}
\caption[]{\label{fig1}The labelling of the cube vertices for the mean field
equations.}
\end{figure}
\clearpage \newpage
\begin{figure}[htb]
\vskip 20.0truecm
\includegraphics{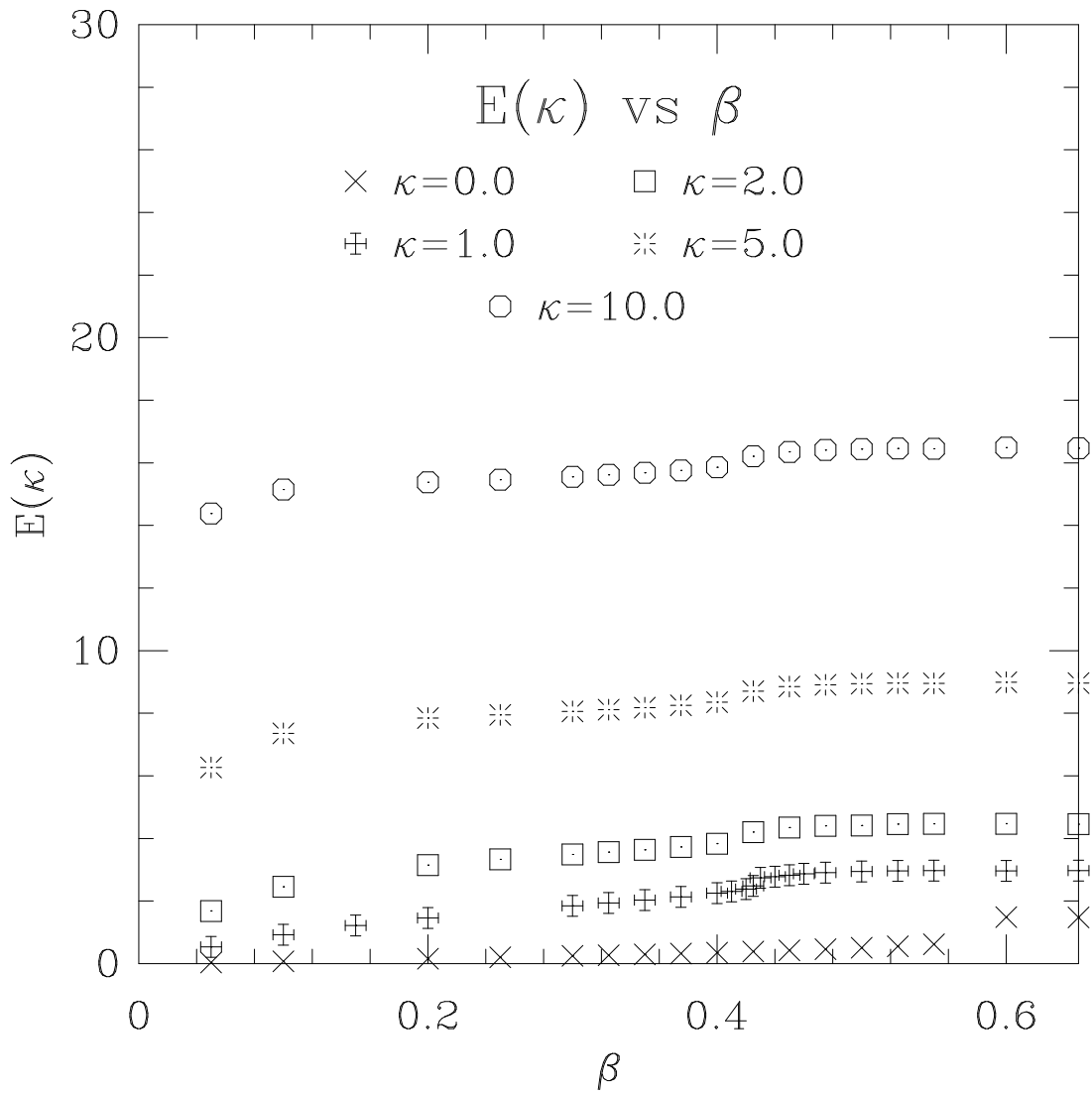}
\caption[]{\label{fig2}The energies for various $\kappa$, all on lattices of size $L=20$.}
\end{figure}
\clearpage \newpage
\begin{figure}[htb]
\vskip 20.0truecm
\includegraphics{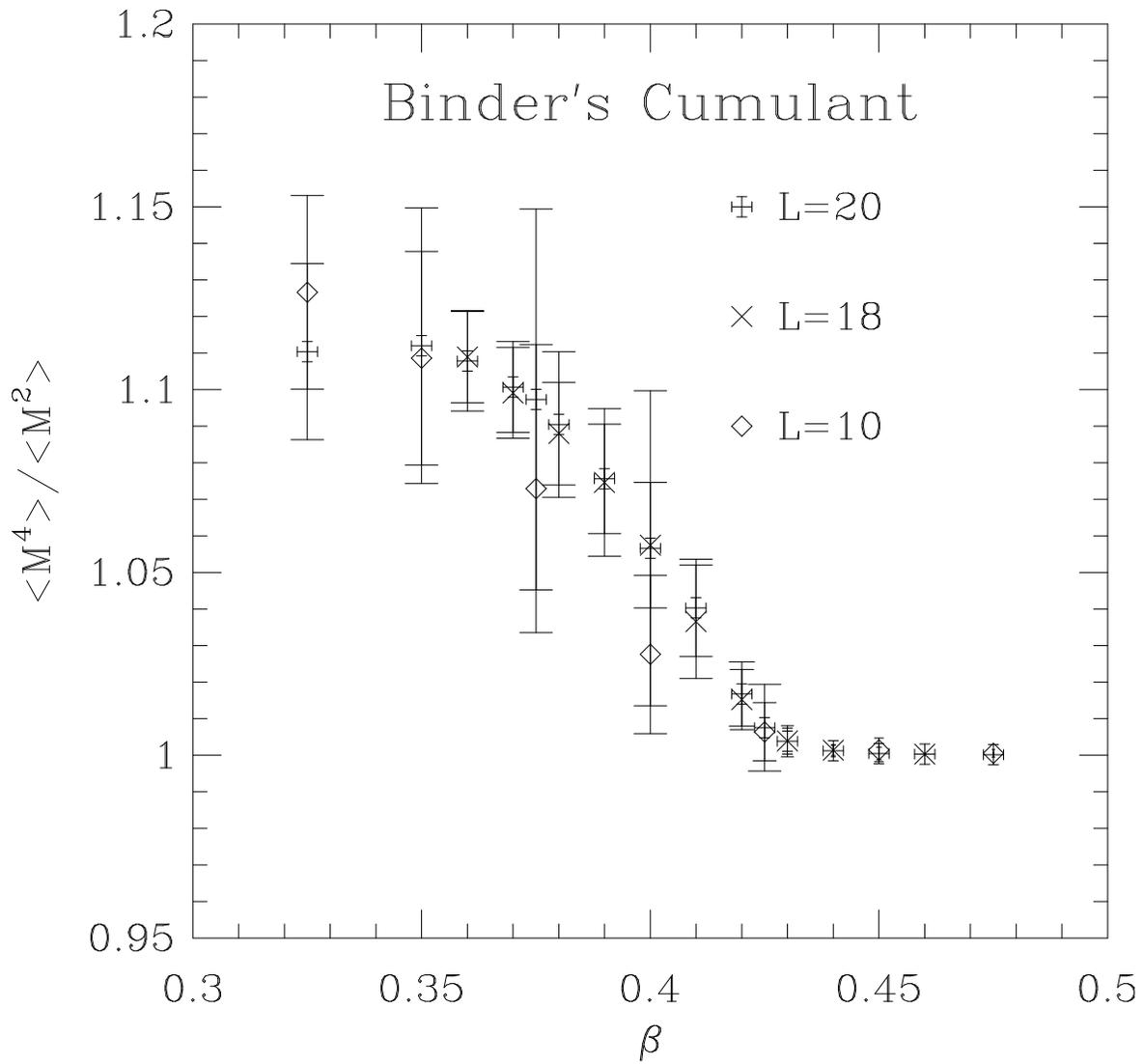}
\caption[]{\label{fig3}Binder's cumulant for $\kappa=1$}
\end{figure}
\clearpage \newpage
\begin{figure}[htb]
\vskip 20.0truecm
\includegraphics{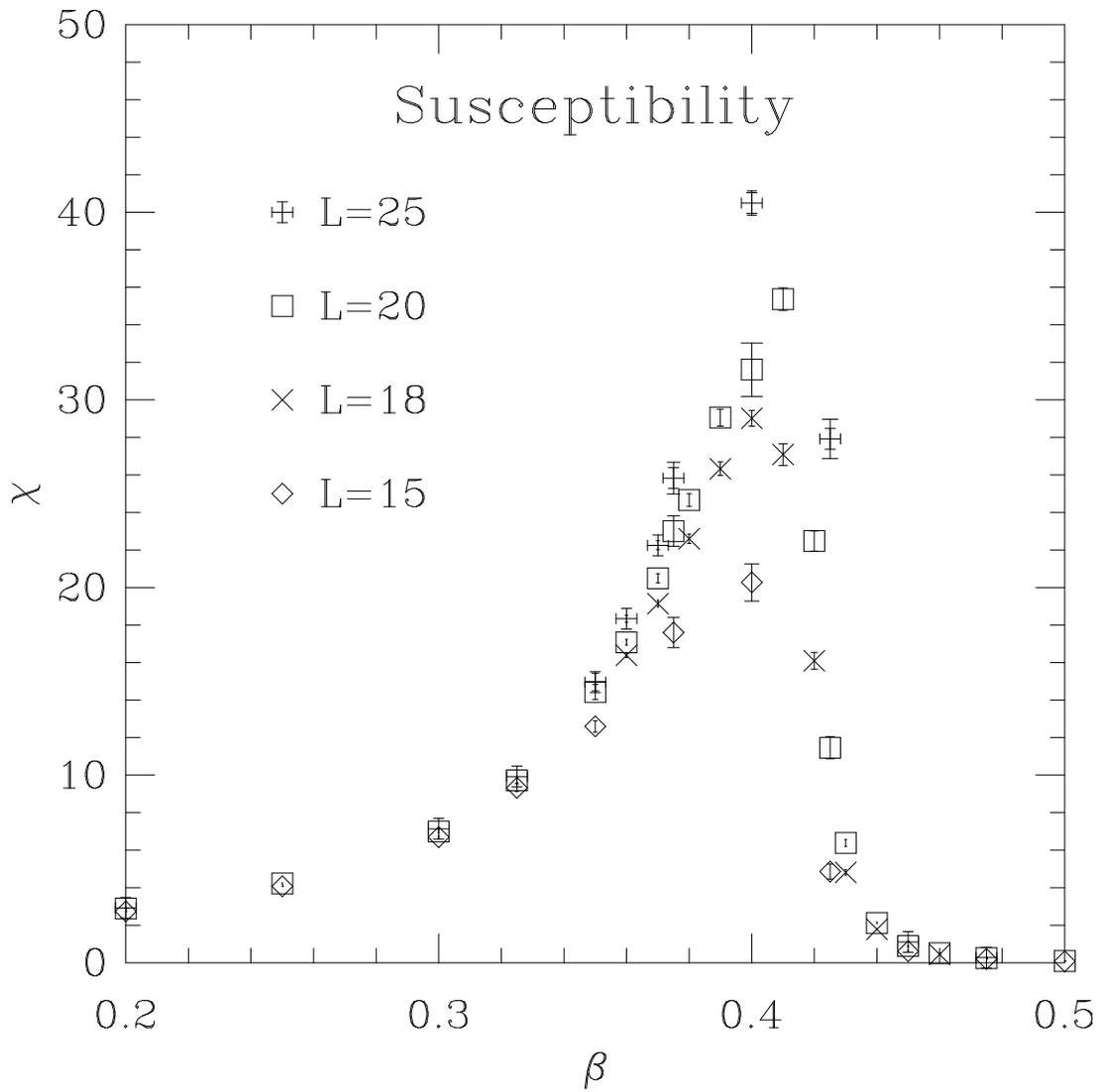}
\caption[]{\label{fig4}The susceptibility for $\kappa=1$}
\end{figure}
\clearpage \newpage
\begin{figure}[htb]
\vskip 20.0truecm
\includegraphics{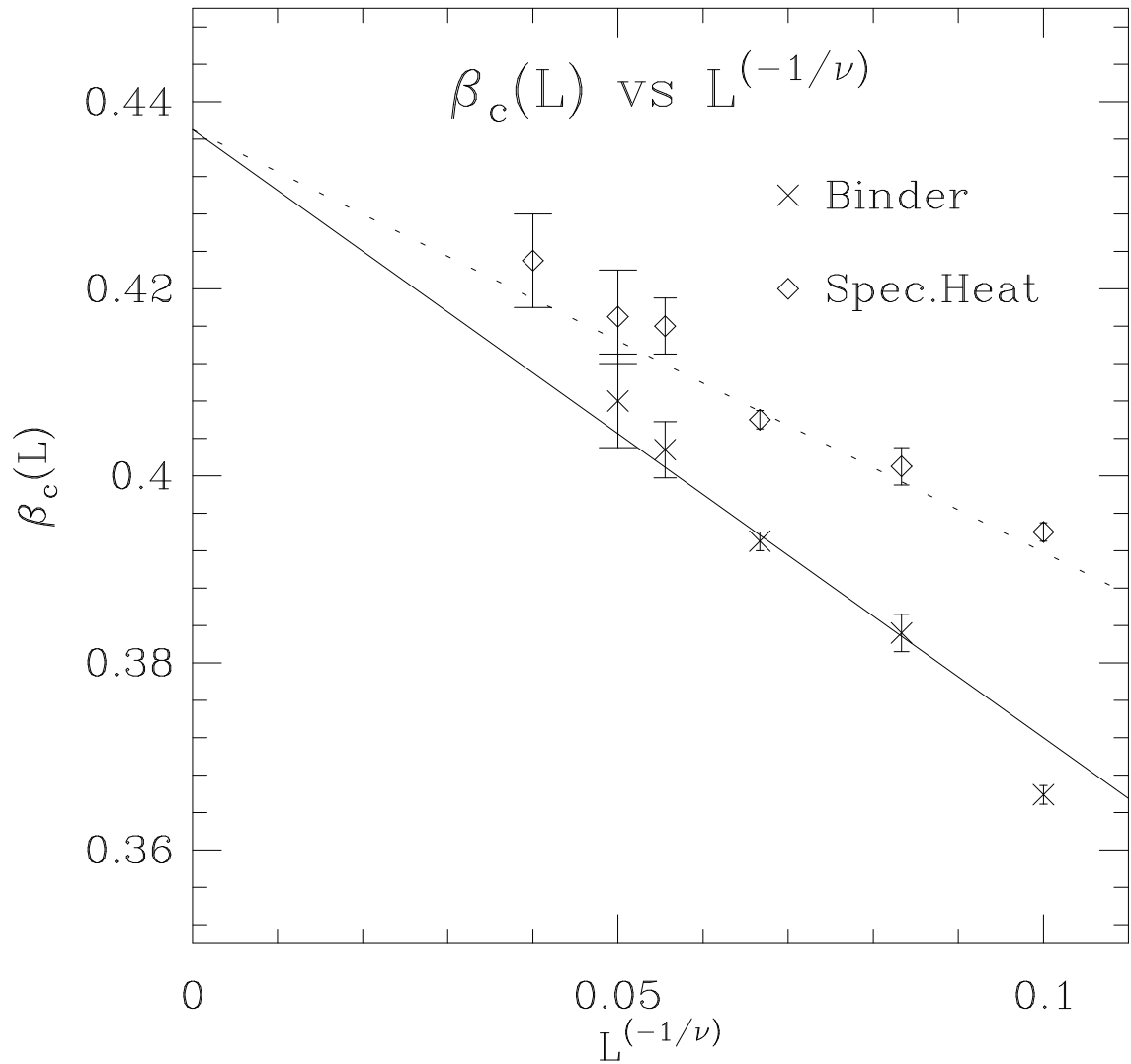}
\caption[]{\label{fig5}The pseudocritical temperature estimated from
the specific heat and Binder's cumulant {\it vs} $L^{-{1 \over \nu}}$ at $\kappa=1$}
\end{figure}
\clearpage \newpage
\begin{figure}[htb]
\vskip 20.0truecm
\includegraphics{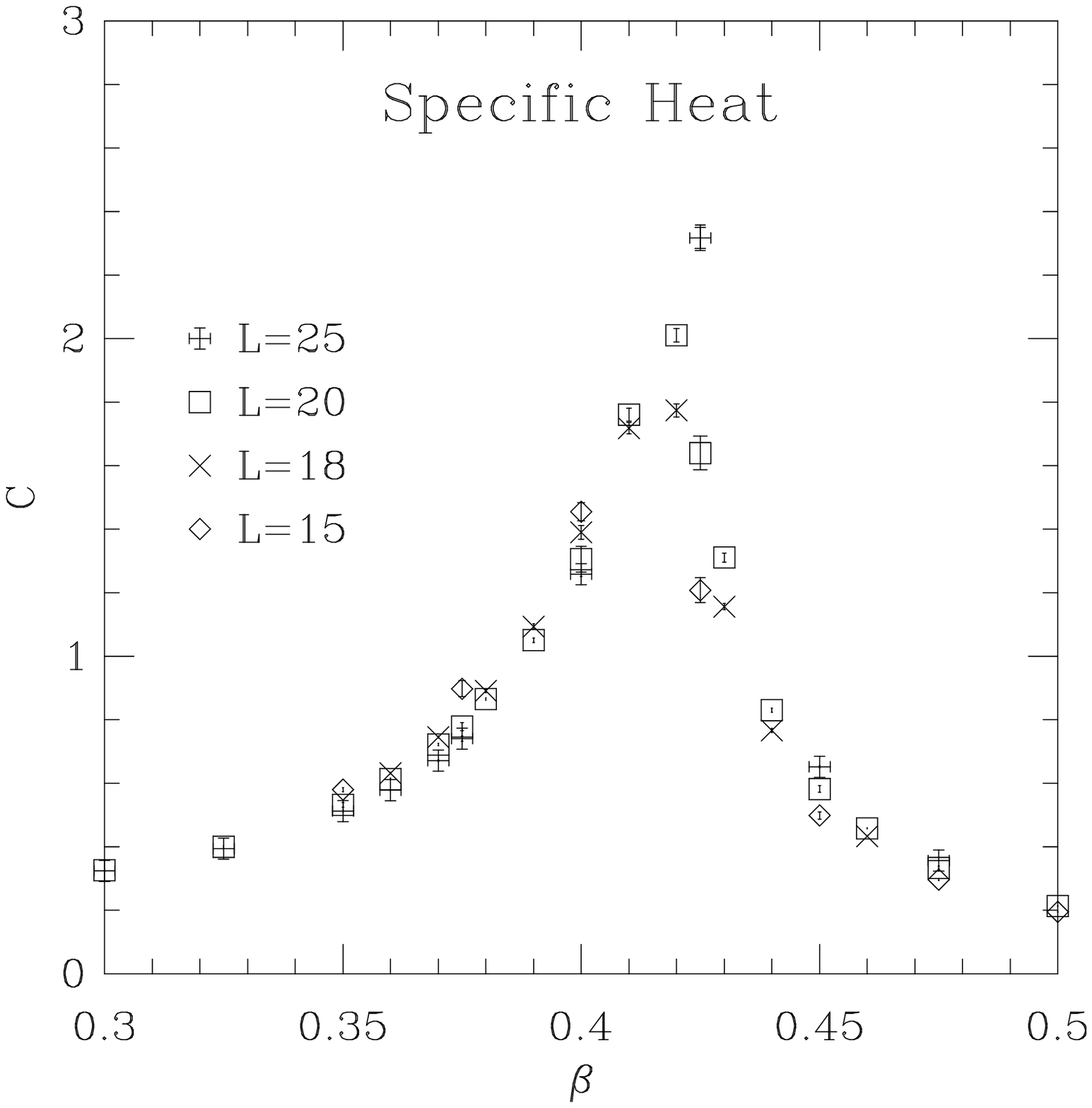}
\caption[]{\label{fig6}The specific heat for $\kappa=1$.}
\end{figure}

\begin{thebibliography}{99}
\bibitem{1} R.V. Ambartzumian, G.S. Sukiasian, G. K. Savvidy
            and K.G. Savvidy, Phys. Lett. {\bf B275} (1992) 99;\\
            G. K. Savvidy and K.G. Savvidy, Int. J. Mod. Phys. {\bf A8} (1993) 3393;\\
            G. K. Savvidy and K.G. Savvidy, Mod. Phys. Lett. {\bf A8} (1993) 2963.
\bibitem{2} S. Catterall, Phys. Lett. {\bf B220} (1989) 207;\\
             C. F. Baillie, D. A. Johnston and R. D. Williams, Nucl. Phys. {\bf B335}
             (1990) 469;\\
             R. Renken and J. Kogut, Nucl. Phys. {\bf B354} (1991) 328;\\
             S. Catterall, Phys. Lett. {\bf B243} (1990) 121;\\
             C. F. Baillie, R. D. Williams, S. M. Catterall and D. A. Johnston,
             Nucl. Phys. {\bf B348} (1991) 543;\\
             S. Catterall, D. Eisenstein, J. Kogut and R. Renken, Nucl. Phys. {\bf B366} (1991) 647;\\
             J. Ambjorn, J. Jurkiewicz, S. Varsted, A. Irback and B. Petersson, Phys. Lett. {\bf B275} (1992) 295;\\
             J. Ambjorn, J. Jurkiewicz, S. Varsted, A. Irback and B. Petersson, Nucl. Phys. {\bf B393} (1993) 571;\\
             M. Bowick, P. Coddington, L. Han, G. Harris and E. Marinari, Nucl. Phys. {\bf B394} (1993) 791.
\bibitem{3} A. Polyakov, Nucl. Phys. {\bf B268} (1986) 406;\\
             H. Kleinert, Phys. Lett. {\bf B174} (1986) 335;\\
             W. Helfrich, J. Phys. {\bf 46} (1985) 1263.
\bibitem{4} C.F. Baillie and D. A. Johnston, Phys. Rev. {\bf D45} (1992) 3326.
\bibitem{5} B. Durhuus and T. Jonsson, Phys. Lett. {\bf B297} (1992) 271.
\bibitem{5a} C.F. Baillie, D.  Espriu and D. Johnston, Phys. Lett. {\bf B305} (1993) 109.
\bibitem{6} C.F. Baillie, A. Irback and W. Janke and D. A. Johnston, Phys. Lett. {\bf B325} (1994) 45.
\bibitem{7} G. K. Savvidy and F.J. Wegner, Nucl. Phys. {\bf B413} (1994) 605;\\
            G. K. Savvidy and K.G. Savvidy, Phys. Lett. {\bf B324} (1994) 72;\\
            G. K. Savvidy, K.G. Savvidy and P.G. Savvidy, ``Dual Statistical Systems and Geometric
            String,'' hep-th/9409031;\\
            G. K. Savvidy and K.G. Savvidy, Phys. Lett. {\bf B337} (1994) 333;\\
            G. K. Savvidy, K.G. Savvidy and F.J. Wegner, Nucl. Phys. {\bf B443} (1995) 565.
\bibitem{8} G. K. Bathas, K. G. Floratos, G. K. Savvidy and K.G. Savvidy, ``Two Dimensional
            and Three Dimensional Spin Systems with Gonihedric Action'', hep-th/9504054.
\bibitem{8a} G. K. Savvidy and K.G. Savvidy, ``Interaction Hierarchy: Gonihedric
String and Quantum
            Gravity,''hep-th/9506184.
\bibitem{8b} R. Pietig and F. Wegner, ``Phase Transition in Lattice Surface System
              with Gonihedric Action'', Heidelberg preprint, December 95. 
\bibitem{9} A. Cappi, P. Colangelo, G. Gonella and A. Maritan, Nucl. Phys. {\bf B370} (1992) 659.
\bibitem{9a} T. Sterling and J. Greensite, Phys. Lett. {\bf B121} (1983) 345;\\
             M. Karowski and H. Thun, Phys. Rev. Lett. {\bf 54} (1985)
2556;\\
             M. Karowski, J. Phys. {\bf A19} (1986) 3375.
\bibitem{10} W. Selke, Phys. Rep. {\bf 170} (1988) 213;\\
             D.P. Landau and K. Binder, Phys. Rev. {\bf B31} (1985) 5946.
\bibitem{11} V. Dotsenko, M. Picco, P. Windey, G. Harris, E. Martinec and E. Marinari,
             ``Self-Avoiding Surfaces in the 3-D Ising model'', hep-th/9504076;\\
             V. Dotsenko, M. Picco, P. Windey, G. Harris, E. Martinec and E. Marinari,
             Phys. Rev. Lett. {\bf 71} (1993) 811.
\bibitem{wh} W. Janke and C. Holm, J. Phys. {\bf A27} (1994) 2553.
\bibitem{esp} M. Baig, D. Espriu, D. Johnston and R.P.K.C. Malmini,
``String Tension in Gonihedric Ising Models'', to appear. 
\end{thebibliography}
\end{document}